\documentstyle[twocolumn,aps]{revtex}
\begin{document}
\wideabs{
\draft
\title{Charge and critical density of strange quark matter}
\author{G. X. Peng,$^{1,2,3}$ H. C. Chiang,$^{2}$
        P. Z. Ning,$^{3}$ and B. S. Zou$^2$}
\address{
         $^1$China Center of Advanced Science and Technology
               (World Laboratory), Beijing 100080, China \\
         $^2$Institute of High Energy Physics,
               Academia Sinica, Beijing 100039, China  \\
         $^3$Department of Physics, Nankai University,
                                Tianjin 300071, China \\
         }
\date{Received 16 November 1998}
\maketitle
\begin{abstract}

The electric charge of strange quark matter is of vital importance
to experiments. A recent investigation shows that strangelets are
most likely highly negatively charged, rather than slightly positively
charged as previously believed. Our present study indicates that
negative charges can indeed lower the critical density, and thus be
favorable to the experimental searches in heavy ion collisions.
However, too much negative charges can make it impossible
to maintain flavor equilibrium. \{S0556-2813(99)02506-6\}

\end{abstract}
\pacs{PACS numbers: 24.85.+p, 12.39.-x, 12.38.Mh, 25.75.-q}
         }

The last 20 years have witnessed an ever-increasing interest in
strange quark matter (SQM) \cite{PengCCAST}. Early in 1970's
\cite{Bodmer}, it had already been known that the appearance of $s$
fraction in two flavor quark droplets provides an additional degree of
freedom, and thus could lower the energy of the system. In 1984,
Witten conjectured that quark matter with strangeness per baryon of
order unity might be bound \cite{Witten}, which has aroused a great
focus of investigations into the stability and detectability of SQM.
In the framework of MIT bag model, Jaffe {\sl et al.}\ find that SQM
is absolutely stable around the normal nuclear density
for a wide range of parameters \cite{Jaffe} while
other authors find a few destability factors \cite{Madsen,Parija}.
On application of the quark mass-density-dependent model, Chakrabarty
{\sl et al.}\ carried out a lot of investigations with significantly
different results \cite{Chakrabarty1,Chakrabarty2}. However, the
investigations by Benvenuto and Lugones give similar results to those
in the bag model \cite{Benvenuto}. A more recent investigation indicates
a link of SQM to the study of quark condensates \cite{PengPRC}.

   The most important way of the territorial searches for SQM is in
the altra-relativistic heavy ion collisions \cite{Greiner}.
Of vital importance to the experimental searches is the electric charge
of SQM. Previously, it is generally believed that SQM is
slightly positively charged \cite{Jaffe}. Contrary to previous findings,
a recent investigation shows that metastable strangelets are most likely
highly negatively charged \cite{Schaffner-Bielich}. This significant
result may, if really the case, have a serious impact on the high
sensitivity searches in heavy ion experiments at the AGS and CERN
facilities. Our present study, in a very simple manner, indicates that
proper negative charges can indeed lower the critical density under which
SQM can not maintain its flavor equilibrium, and thus no longer exist.

  Following Ref.\ \cite{Jaffe}, We assume the SQM to be a Fermi gas
mixture of $u$, $d$, $s$ quarks and electrons with chemical equilibrium
maintained by the weak interactions:
$
 d, s \leftrightarrow u+e+\overline{\nu}_{e}, \, \,
                     s+u \leftrightarrow u+d.
$

   At zero temperature, the thermodynamic potential density of the
species $i$ is
\begin{eqnarray}          \label{ztpd}
  \Omega_{i}
 = - \frac{g_{i}}{48\pi^2}
 &\left[\mu_{i}(2\mu_{i}^2-5m_{i}^2)\sqrt{\mu_{i}^2-m_{i}^2}\right.
                                            \nonumber      \\
 \mbox{}&
 \left.+3m_{i}^{4}\ln\frac{\mu_{i}+\sqrt{\mu_{i}^{2}-m_{i}^{2}}}{m_{i}}
   \right],
\end{eqnarray}
where $i=u, d, s, e$ (ignore the contribution from neutrinos);
$g_i$ is the degeneracy factor with values $6$ and $2$, respectively
for quarks and electrons.

   The corresponding particle number density is
\begin{equation}
  n_i=-\frac{\partial\Omega_i}{\partial\mu_i}
     =\frac{g_i}{6\pi^2}(\mu_i^2-m_i^2)^{3/2}.
\end{equation}

For a given baryon number density $n_b$ and electric charge density
$Q$, the chemical potentials $\mu_u$, $\mu_d$, $\mu_s$, and $\mu_e$ 
are determined by the following equations 
\begin{eqnarray}
   &  \mu_d  = \mu_s \equiv \mu,  &    \label{eqmu1}     \\
   & \mu_u + \mu_e = \mu,         &                      \\
   & \frac{1}{3} (n_u + n_d + n_s) = n_b, &              \\
   & \frac{2}{3}n_u-\frac{1}{3}n_d-\frac{1}{3}n_s-n_e = Q.
                                      \label{eqmu4}   &
\end{eqnarray}

   The last two equations are equivalent to
\begin{eqnarray}
          & n_u-n_e=n_b+Q,                  &    \\
          & n_d+n_s+n_e=2n_b-Q.             &
\end{eqnarray}

   We therefore define a function of $\mu_e$
\begin{eqnarray}
  F(\mu_e) &=& \pi^2(n_d+n_s+n_e)-\pi^2(2n_b-Q)          \\
           &=& (\mu^2-m_d^2)^{3/2}+(\mu^2-m_s^2)^{3/2} \nonumber \\
           & & +\frac{1}{3}(\mu_e^2-m_e^2)^{3/2}-\pi^2 (2n_b-Q),
\end{eqnarray}
where
\begin{equation}
  \mu=\mu_e+\sqrt{m_u^2+[\pi^2 (n_b+Q)
           +\frac{1}{3}(\mu_e^2-m_e^2)^{3/2}]^{2/3}}.
\end{equation}

  Because $m_s>m_{u,d}$, the equation  $ F(\mu_e)=0 $
for $\mu_e$ has solution if and only if
\begin{eqnarray}
    &     \mu   \geq  m_s, & \\
    &  F(\mu_e)  \leq  0.  &
\end{eqnarray}

   At the critical density $n_c$ of baryon number, the equality signs
in the above two inequalities should be taken. So we can easily find the
equation
\begin{eqnarray}  
 &   \left\{   m_s-\sqrt{m_u^2+\left[
      3\pi^2n_c-\left(m_s^2-m_d^2\right)^{3/2}
                    \right]^{2/3}}
     \right\}^2          &    \nonumber \\
 & \ \  =m_e^2+\left\{3\left[\pi^2(2n_c-Q)
               -(m_s^2-m_d^2)^{3/2}
             \right]\right\}^{2/3}  &   \label{eqnc}
\end{eqnarray}
which determines how the critical density depends on charge density,
namely, $n_c=f(Q)$.

 At $n_c$, the strangeness fraction becomes zero.
 When the density decreases further, the equation group
 (\ref{eqmu1}--\ref{eqmu4}) which determines the  configuration of
 the system has no solution. This indicates that $n_c$ is the lowest
 density for the possible existence of SQM.

It is obviously difficult to get an explicit expression for the
function $f(Q)$.  However, the inverse function  $f^{-1}(n_c)$ can be
easily obtained from Eq.\ (\ref{eqnc}):
\begin{eqnarray}    
 Q &=& 2n_c-\frac{1}{\pi^2}(m_s^2-m_d^2)^{3/2}  \nonumber \\
   & & -\frac{1}{3\pi^2}
       \left[\left(m_s- \mu^\prime                                                                         
             \right)^2-m_e^2
       \right]^{3/2},              \label{qexp}               
\end{eqnarray}
where
\begin{equation}
   \mu^\prime \equiv \sqrt{m_u^2+\left[ 3\pi^2 n_c-\left(m_s^2-m_d^2
                                            \right)^{3/2}
                           \right]^{2/3}}.
\end{equation}

In order to make the inverse square root meaningful, we must require
\begin{equation}
  \left(m_s^2-m_d^2\right)^{3/2} \leq  3\pi^2 n_c.
\end{equation}

When taking the equality sign, and then solving the corresponding
equation, we obtain the minimum critical density $n_{cmin}$.
Substituting into Eq.\ (\ref{qexp}) gives
\begin{equation}
Q_{\text{min}}=-n_{\text{cmin}}-\frac{1}{3\pi^2} \left[
                        (m_s^\prime-m_u^\prime)^2-m_e^2
                                   \right]^{3/2},
\end{equation}
where $m_s^\prime$\ and $m_u^\prime$\ are, respectively, the masses of
$s$ quarks and $u$ quarks corresponding to the density $n_{cmin}$.

If the charge density $Q$ is more negative than $Q_{\text{min}}$, there
is no critical density corresponding to it. Therefore, $Q_{\text{min}}$
is the most negative charge density of SQM.

To perform a concrete calculation of the critical density corresponding
to $Q>Q_{\text{min}}$, we must stop here for a little while
to discuss the consideration of quark confinement.

Let's schematically write the QCD Hamiltonian density as
\begin{equation}      \label{hqcd}
   H_{\text{QCD}} = H_k + \bar{q}M_0q + H_{\text{I}},
\end{equation}
where $H_k$ is the kinetic term, $H_{\text{I}}$ is the strong interaction
terms between quarks, $M_0$ is the current mass matrix with diagonal
elements $m_{u0}$, $m_{d0}$, and $m_{s0}$.

In the mass-density-dependent model, the strong interaction between
quarks is mimicked by the proper dependence of quark masses on density.
It is equivalent to expressing the  Hamiltonian density as
\begin{equation}      \label{heff}
   H_{\text{equiv}} = H_k + \bar{q}Mq
\end{equation}
which is the same with that of free quarks except that the current mass
matrix $M_0$ is replaced by an equivalent mass matrix $M$ with diagonal
elements $m_q$ $(q=u,d,s)$. From the equality
\begin{equation}      \label{hqcdheff}
  \langle{n_b}|H_{\text{equiv}}|n_b\rangle
  =\langle{n_b}|H_{\text{QCD}}|n_b\rangle,
\end{equation}
where $|n_b\rangle$\ is the state vector with baryon number density $n_b$,
we have
\begin{eqnarray}
     m_q &=& m_{q0} + \frac{\langle{n_b}|H_{\text{I}}|n_b\rangle}
           {\sum\limits_{q}\langle{n_b}|\bar{q}q|n_b\rangle}
                              \label{mi} \\
       &\equiv& m_{q0} + m_{\text{I}}.    \label{mmi}
\end{eqnarray}

    The physical meaning of Eq.\ (\ref{mmi}) is clear: the first
term on the right is the original mass, i.e., the quark current mass,
while the second term $m_{\text{I}}$ is the interacting part mimicking
the strong interaction between quarks. Obviously, $m_{\text{I}}$ is
flavor-independent and density-dependent.  Because the characteristic
of strong interactions is the confinement of quarks, we should require
$\lim_{n_b\rightarrow 0}m_{\text{I}}=\infty$. A popularly used
parametrization form is
\cite{Chakrabarty1,Chakrabarty2,Benvenuto}
\begin{equation}    \label{mq}
   m_q = m_{q0} + \frac{B}{3n_b},
\end{equation}
where $B$ is a fixed constant determined by stability arguments.

Because we are interested in the true ground state of SQM which might
pass into a detector, we must require, at zero pressure, the energy
per baryon is less than $M_{^{56}Fe}/56=930$ MeV for SQM and greater than
930 MeV for two-flavor quark matter (where $M_{^{56}Fe}$ is the mass of
$^{56}Fe$) \cite{Jaffe}. A little later, we will see that the negative
charge density is not permitted to shift too far away from zero.
We thus use the same method with that in Ref.\ \cite{Benvenuto}
to estimate the absolute stability range for the paremeter $B$
(the authors of Ref.\ \cite{Benvenuto} rename it $C$), which gives
(69, 110) MeV$\cdot$fm$^{-3}$.
Taking the modest values $B=80$ MeV$\cdot$fm$^{-3}$,
$m_{u0}=m_{d0}=7.5$ MeV, and $m_{s0}=100$ MeV, we can calculate the
critical density by substituting Eq.\ (\ref{mq}) into Eq.\ (\ref{eqnc})
or (\ref{qexp}). The result is
plotted in Fig.\ 1 with a solid line. It is obvious that $n_c$ is an
increasing function of $Q$. Negative charges correspond to smaller
critical densities. However, when $Q$ decreases further toward negative
direction, there will be no $n_c$ corresponding to it because SQM can
not maintain its flavor equilibrium at that case.
Therefore, we conclude that proper negative charges can lower the critical
density, and thus be favorable to the experimental searches.

To explore if the above conclusion depends on mass formulas used,
we now adopt the following expression:
\begin{eqnarray} \label{mq2}
   m_q = m_{q0} + \frac{D}{n_b^2},
\end{eqnarray}
obtained for reasons of symmetry (particle-antiparticle
conjugation, Lorentz transformations). Eq.\ (\ref{mq2}) is obviously
in consistency with the properties of the mass (invariant under
particle-antiparticle conjugation, Lorentz-scalar; here the density
is replaced by square of the four-current $j_{\mu}j^{\mu}$).
The constant $D$ should be in the range of (113--130 MeV)$^7$ according
to the method in Ref. \cite{Benvenuto}.
Taking $D=(120$ MeV)$^7$ and the same values of $m_{q0}$ with the above, 
we plot the corresponding critical density as a function of the charge
density in Fig.\ 1 with a dashed line.

We also perform a calculation in the bag model where the quark
confinement is mimicked by adding to the energy density expression an
extra constant which is interpreted as the vacuum pressure, while the
quark masses are density-independent:
\begin{eqnarray}
        &    m_{u,d} = 7.5\  \mbox{MeV}, &  \\
        &    m_s     = 100\ \mbox{MeV}.  &
\end{eqnarray}
The corresponding result is given in Fig.\ 1 with a dotted line.

    The shapes of the three lines in Fig.\ 1 are obviously similar to
each other and so lead to the same conclusion as mentioned in the above.
The significant difference is that the line corresponding to bag model
calculation is much lower, or in other words, the charge density of
stable bulk SQM in the bag model is nearly not allowed to be negative.
This is due to the complete ignorance of  the strong interactions
in the bag model calculation. At high densities, they asymptotically
become identical as expected.

  It should be pointed out that our method of reasoning is different
from what given by Schaffner-Bielich {\sl et al}.\
\cite{Schaffner-Bielich}. There the investigation was concerned with
how small metastable strangelets (including explicit finite size
corrections) look like and might decay for different scales of lifetime.
None of the candidates found was being absolutely stable, but this fact,
of course, depends on the bag parameter being employed. There it was
not assumed, as done here, that the strangelets are in weak chemical
equilibrium, but it was investigated how the metastable candidates
might look like if they are assumed to be stable against strong hadronic
decay and subsequently against weak hadronic decay. These decay modes
drive the more stable candidates to negative charges.
Here we have assumed 
that the pieces of SQM are in perfect weak chemical equilibrium and 
considered as having the possibility of absolute stability. Within
the standard MIT approach, these pieces of stable bulk SQM would have a
slightly positive charge. It is shown here, 
with the quark mass-density-dependent model,
that  the overall charge in fact might be
slightly negative. Naturally, the present treatment is very simple,
so the concrete values should not be taken very seriously, and
further studies are needed. 

  In conclusion, the present investigation is in favor of the viewpoint
that {\em proper} negative charges can lower the critical density and
consequently be favorable to the experimental searches for strange
quark matter in ultra-relativistic heavy ion collisions.

\begin{center}
\section*{          ACKNOWLEDGMENTS }
\end{center}

The authors would like to thank the partial financial support by
National Natural Science Foundation of China under Grant Nos.
19675044 and 19875026.

\begin{figure}
\caption{Critical density $n_c$ vs electric charge density $Q$.
The three lines correspond to different mass formulae
(for details, see text).
We can see that $n_c$ is an increasing function of $Q$, i.e.,
negative charges can lower the critical density.
The points marked with ``$\bigcirc$`` correspond to
the most negative charge density permitted.
       }
\label{fignm}
\end{figure}


\begin{references}
\bibitem{PengCCAST}
G.X. Peng, P.Z. Ning, and S.J. Luo, 
    CCAST-WL Workshop Series, {\bf 57}, 1 (1996).


\bibitem{Bodmer}
 A.R. Bodmer,
    Phys.\ Rev.\ D {\bf 4}, 1601 (1971);
 S.A. Chin and A.K. Kerman,
    Phys. Rev. Lett. {\bf 43}, 1292 (1979).

\bibitem{Witten}
 E. Witten,
    Phys.\ Rev.\ D {\bf 30}, 272 (1984).

\bibitem{Jaffe}
 E. Farhi and R L. Jaffe,
    Phys.\ Rev.\ D {\bf 30}, 2379 (1984);
 M.S. Berger and R.L. Jaffe,
    Phys.\ Rev.\ C {\bf 35}, 213 (1987);
 E.P. Gilson and R.L. Jaffe,
    Phys.\ Rev.\ Lett.\ {\bf 71}, 332 (1993).

\bibitem{Madsen}
 Jes Madsen,
    Phys. Rev. Lett. {\bf 61}, 2909 (1993);
    Phys. Rev. D {\bf 47}, 5156 (1993);
                 {\bf 50}, 3328 (1994);
 Jes Madsen, Dan M. Jensen, and Michael B. Christiansen, 
    Phys. Rev. C {\bf 53}, 1883 (1996).

\bibitem{Parija}
 B.C. Parija, 1993,
     Phys. Rev. C {\bf 48}, 2483 (1993);
                  {\bf 51}, 1473 (1995).

\bibitem{Chakrabarty1}
 S. Chakrabarty, S. Raha, and B. Sinha,
    Phys.\ Lett.\ B {\bf 229}, 112 (1989).

\bibitem{Chakrabarty2}
 S. Chakrabaty,
    Phys. Rev. D {\bf 43}, 627 (1991);
                 {\bf 48}, 1409 (1993);
                 {\bf 54}, 1306 (1996).

\bibitem{Benvenuto}
 O.G. Benvenuto and G. Lugones,
    Phys.\ Rev.\ D {\bf 51}, 1989 (1995);
 G. Lugones and O.G. Benvenuto,
    {\sl ibid.}\ {\bf 52}, 1276 (1995).

\bibitem{PengPRC}
 G.X. Peng, P.Z. Ning, and H.C. Chiang,
    Phys. Rev. C {\bf 56}, 491 (1997).

\bibitem{Greiner}
 C. Greiner, P. Kock, and H. St\"{o}cker,
    Phys.\ Rev.\ Lett.\ {\bf 58}, 1825 (1987);
 C. Greiner, D. H. Rischke, H. St\"{o}cker, and P. Koch,
    Phys.\ Rev.\ D {\bf 38}, 2797 (1988).
 C. Greiner, P. Koch, and H. St\"{o}cker,
   {\sl ibid.}\ {\bf 44}, 3517 (1991).

\bibitem{Schaffner-Bielich}
 J\"{u}rgen Schaffner-Bielich, Carsten Greiner,
           Alexander Diener, and Horst St\"{o}cker,
    Phys. Rev. D {\bf 55}, 3038 (1997).
 
\end{references}
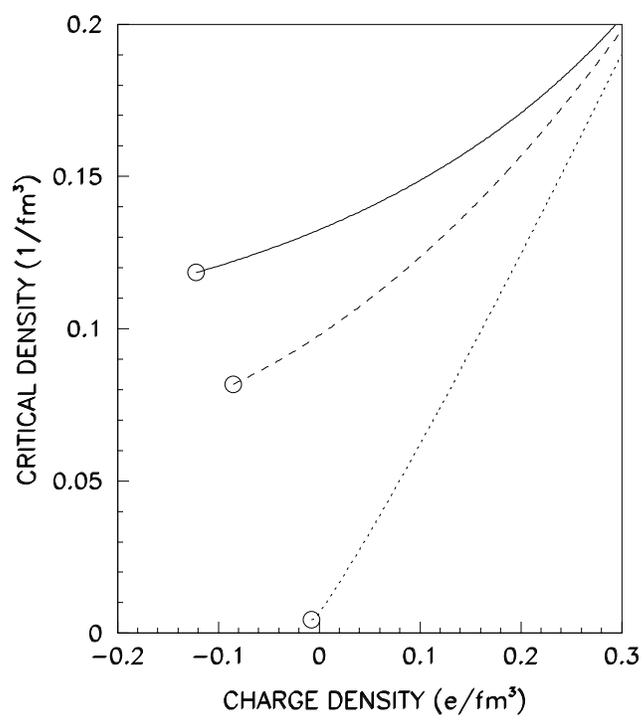
\end{document}